\documentstyle[epsfig]{mn}

\title{Detailed comparison of the structures and kinematics of simulated and observed barred galaxies }
\author[J. O'Neill and J. Dubinski]
       {J.K. O'Neill$^{1,2}$\thanks{E-mail: oneill@cita.utoronto.ca} and
	    John Dubinski$^{1,2}$\thanks{E-mail: dubinski@cita.utoronto.ca}\\
$^1$Canadian Institute for Theoretical Astrophysics, 
University of Toronto, Canada \\
$^2$Department of Astronomy \& Astrophysics, University of Toronto, Toronto, Canada}
\date{Received: \ \ \ Accepted: }

\pagerange{\pageref{firstpage}--\pageref{lastpage}}
\pubyear{2002}

\begin{document}

\maketitle

\label{firstpage}

\begin{abstract}

We examine the observable properties of simulated barred galaxies including
radial mass profiles, edge-on structure and kinematics, bar lengths 
and pattern speed evolution for detailed comparison to real systems.
We have run several simulations in which bars are
created through inherent instabilities in self-consistent
simulations of a realistic disc+halo galaxy model with a disc-dominated, flat
rotation curve. These simulations were run at high
(N=20M particles) and low (N=500K) resolution to test numerical
convergence.  We determine the pattern speeds in simulations
directly from the phase angle of the bar versus time
and the Tremaine-Weinberg method. 
Fundamental dynamics do not change between the high and low
resolution, suggesting that convergence has been reached in this
case. We find the higher resolution is needed to simulate structural and kinematic
properties accurately. 
The edge-on view of the higher-resolution system clearly shows the 
bending instability and formation
of a peanut-shaped bulge.
We determined bar lengths by different means to determine the 
simulated bar is fast, with a corotation to bar length ratio under $1.5$.
Simulated bars in these models form with pattern
speeds slower than those observed and slow down during their 
evolution. Dynamical friction between the bar and dark halo is
responsible for this deceleration, as revealed by the transfer of
angular momentum between the disc and the halo. However, even though the 
pattern speed is reduced at later times, the instantaneous scale length of 
the disc has grown sufficiently for the bar motion to agree with many 
observations.  
By using a different
model and simulation technique than other authors, we are able to 
compare the robustness of these methods.  An animation of the face-on and 
edge-on views of the 20M particle simulation is available at http://www.astro.utoronto.ca/~oneill .

\end{abstract}


\section{Introduction}

The dichotomy between the regular and the barred spirals in
Hubble's original galaxy classification scheme is a long standing
problem in galaxy evolution and dynamics.  It is unclear what causes a 
spiral galaxy to become barred or not, or more quantitatively,  
why barred galaxies represent about 30 to 70 per cent (deVaucouleurs 1963; 
including weak bars, Sellwood \& Wilkinson 1993; viewed in IR, Eskridge 
et al. 2000) of nearby galaxies.  This picture becomes even more confusing 
at high redshifts, where Abraham et
al. (1999) have found the fraction to drop beyond a redshift of 0.5.  
Bars undoubtedly form from dynamical
instabilities inherent in self-gravitating axisymmetric discs.
Early simulations (e.g. Ostriker \& Peebles 1973) have shown that
purely self-gravitating discs are instantly susceptible to a bar
instability, while the addition of a surrounding spheroidal
gravitational potential can prevent bar formation; a more recent study 
suggests this is merely a slow down of formation, and the bar is 
actually enhanced at later stages (Athanassoula 2002a).  The
surrounding stabilizing spheroids can be identified with the dark
haloes believed to surround spirals and are responsible for
the nearly flat rotation curves of all galaxies.  Cold dark matter
models of cosmology predict that most spirals are embedded within
nearly isothermal haloes (Dubinski \& Carlberg 1991; Navarro, 
Frenk \& White 1996, 1997).  
It is therefore useful to examine the
consequences of dark haloes on the formation and evolution of barred galaxies
to seek consistency with the prevailing world model and the
observable universe.

The pattern speed of the bar is an important although difficult to
measure indicator of disc dynamics and dark halo structure.
Observations of early-type barred galaxies (Merrifield \& Kuijken 1995 
(M\&K);
Gerssen et al. 1999) through the application of the
Tremaine-Weinberg method (Tremaine \& Weinberg 1984b) (T\&W) have shown 
the pattern speed to be
`fast', since the bars end near corotation.
N-body models of disc dynamics with static background haloes
(summarized in Sellwood 1981) have shown the pattern speed also to be
fast.  However, 
soon afterward it was
recognized (Tremaine \& Weinberg 1984a; Weinberg 1985) that rotating 
bars would spin down significantly due to dynamical friction from the 
dark halo. 
More recent simulations with
self-consistent disc and halo distributions clearly showed this
effect:  angular momentum is transferred from the bar to the
halo through torques due to gravitational wakes in the halo, which
result in surprisingly low pattern speeds for simulated bars
(Debattista \& Sellwood 2000 (D\&S); Misiriotis \& Athanassoula 2000).
D\&S conducted parametric studies by
varying the halo to disc mass ratio of their models and found that
a maximal disc yields the least pattern speed slow down. They
argued that the observational evidence of fast bars implies 
a maximum disc in all barred galaxies.

Bar structure and dynamics can also be examined in edge-on
systems.  The thin bars that form in discs are subject to a
buckling instability which causes a bar to bend vertically and
thicken into a bulge-like object within a few dynamical times,
as was shown numerically by
Raha et al. (1991), Pfenniger \& Friedli (1991), 
and Combes \& Sanders (1981).  Kuijken \&
Merrifield (1995) (K\&M) have shown that the kinematics of peanut-shaped 
bulges of NGC 5746 and NGC 5965 are
consistent with
orbits in a bar potential, the theory of which was confirmed by 
 Bureau \& Athanassoula (1999) (B\&A).
The observational test described by K\&M can
also be applied to
simulations, which can be viewed at any angle for more complete
results.

Through B and I band observations of 15 galaxies, Elmegreen \& 
Elmegreen (1985) have shown that early- (SB0, SBa) and late-type bars
have significantly different properties.  Although resonance
positions and bar lengths are hard to quantify observationally,
there is evidence to suggest that early-type bars are
longer, out to the corotation radius, have a flat surface brightness
profile along their length, and end beyond the turnover radius of
the rotation curve.  The late-type bars on the other hand are much
shorter, with exponential surface brightness profiles.  

In this paper,  we revisit the problem of the bar and buckling
instabilities in self-consistent disc galaxy models with live
haloes. Our goal is to study the pattern speed evolution of
numerical bars as well as quantify the structural and kinematic
properties for comparison with observed face-on and edge-on barred
galaxies. We have taken care to re-scale our models to
observed systems for comparison.
We extend previous work by introducing a new, more realistic mass model and
going to much higher resolution with a simulation containing N=20M
particles,  a factor of 20 times larger than most work on the
subject. Our motivation for using high resolution is to examine the
numerical convergence of results, since
subtleties of dynamical interactions between discs and haloes
may not be captured correctly even with haloes with more than 
$\approx
1$M particles (e.g. Weinberg 2001).  The disc of our large
simulation contains 10M particles, taking us well out of the
regime where disc self-heating contaminates results. We are
confident that the dynamics of these models reliably
represents the gravitational behaviour of real galaxies. There are also
disagreements between various simulations of bars in the
literature which use different N-body codes and initial
conditions.   We attempt here to achieve convergence by
comparing three of our simulations with those of other groups, as well
as to observations, by
examining the bar structure, pattern speed, and edge-on
kinematics.

\section{Mass Models}

We simulate bars by setting up an initially axisymmetric
system including a disc and dark halo which is formally in
equilibrium but unstable.  Models with rotation curves which 
are disc-dominated, approaching the maximal disc as defined by 
van Albada \& Sancisi (1986), usually
form a bar within a few dynamical times, and the Ostriker \& Peebles
(1973) empirical criterion is still a useful indicator of the
instability.  

We use the method of Kuijken \&
Dubinski (KD) (1995) to generate self-consistent disc+halo models
with a nearly flat rotation curve.
We consider models with disc-dominated rotation curves with and without
a central bulge.  
The KD models are generated from a distribution function (DF) that is the
sum of up to 3 functions: a three-integral disk DF, a bulge DF modelled as
King model with an energy cut-off $E < 0$ and a halo DF that is a
flattened King model DF (or lowered Evans model) with the usual 
truncation at a tidal radius at $E=0$.   
We examine models with and without a bulge as described below.
The disk DF is a
function of E, $z$-angular momentum, $L_z$ and a third approximate integral, the
$z$ energy, $E_z = \dot{z}^2/2 + \Phi(z)$.  The disk DF is constructed
assuming a exponential radial surface density profile 
and an approximately $\rm{sech}^2 z$
edge-on profile with fixed vertical scale length $z_d$.  The disk
squared radial velocity dispersion, $\sigma_R^2$ is assumed 
to decline exponentially with the same scale-length as the disk like real
galaxies.  We can generate N-body realizations of these models and they are
formally in equilibrium with an initial virial ratio $2T/W=-1.0$.

The bulgeless model is nearly a
formal maximal disc model, with the rotation curve rising to a
roughly flat profile within two scale lengths (Figure 1).
Although the halo DF is a King model, the mass profile does not
have the usual core since the disk mass dominates the centre.
The halo density profile has a mild cusp with $\rho \propto r^{-0.7}$
to within 0.1 disk exponential scale lengths similar to an NFW profile.
Whether or not the haloes of spiral galaxies have a central core is
still controversial but our model halo here is consistent with rotation
curve decompositions (e.g. Kent 1986).
The disc is in formal equilibrium with a Toomre $Q\sim$1.1 
measured at $R$=2.0$R\rm _d$ and is approximately constant in the range
$1.0 < R/R_d < 4.0$ rising to higher values beyond these limits.
While the model is stable against axisymmetric perturbations it is
inherently unstable to bar formation because of the dominance of the disk
in the rotation curve.  The model is scaled such 
that $G$=$R_d$=$v_{max}$=1, and the total
disc+halo mass is about 12 (disc mass of 1.98 and halo mass of 10.57). 
We realize these models as N-body systems at high resolution with
N=20M particles (10M each in the disc and the
halo) and low resolution with N=500K particles (250K each in the disc and
halo).   
Thus, all quoted simulation lengths are in terms of
the exponential disc length, and all velocities are fractions of
the maximum.  
We also consider a N=500K model with a small compact bulge of mass $M=0.32$
with halo and disk mass profiles essentially unchanged.  The Toomre $Q$ 
also remains comparable at $\sim 1.06$.
The code and parameters for generating these models are available by
request.
Our models
differ from D\&S who use a
Kuz'min-Toomre disc with a vertical Gaussian profile and a lowered
polytrope distribution function (DF) for the halo. The recent 
publication by
Athanassoula \& Misiriotis (2002) (A\&M) contains models
based on Hernquist's (1993) method.
\begin{figure*}
\centering \epsfig{figure=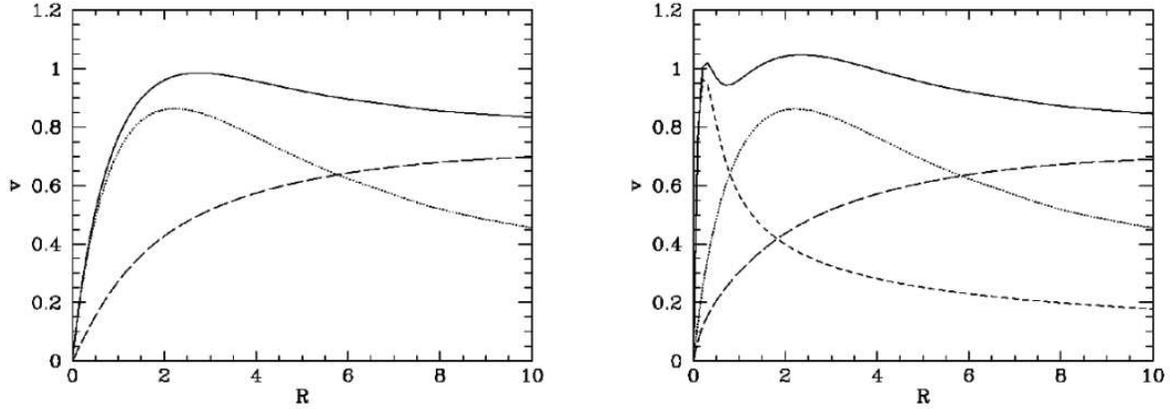,width=6.5in,angle=0}
\caption{Rotation curves for disc+halo (left) and disc+bulge+halo (right) models in simulation units.  The contribution of the disc is shown in dots, the halo in long dashes, the bulge in short dashes and the total as a solid line. R is measured in 
exponential disk scale lengths.}
\end{figure*}  
\begin{figure}
\centering \epsfig{figure=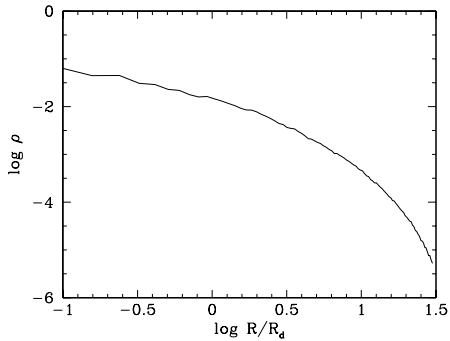,width=6.5cm,angle=-90}
\caption{Initial halo density profile:  a King model halo with a very 
small core; the profile is similar to an NFW profile with a -0.7 power law 
slope until at least $R\sim$0.1}
\end{figure}

As well as specifying the following results in the original
simulation units, the appropriate lengths and times are scaled
to match the SBa galaxy NGC 4596 for easier comparison with real
galaxies. We use the scale length of $3.2 \rm{\ kpc}$ and
rotational velocity of $120\rm{\ km\ s}^{-1}$ given by Gerssen et
al. (1999).  The $\rm{sech}^2$ scale height for the disc is 0.1
simulation units or 320 pc.  For
these scales the unit of time is 26.3 Myr. For the runs, we use a
time step of $\Delta t=0.05$ (1.32 Myr) for 10,000 time steps, 
for a total $t$=500 (13.2 Gyr) or approximately the
Hubble time.  We soften gravitational forces using a Plummer law
with softening length $\epsilon=0.0125$ (40 pc) for the halo and
$\epsilon=0.005$ (16 pc) for the disc.  All runs are executed
with a parallel tree code
(Dubinski 1996) running on a PC cluster or a 32 processor Compaq GS320. 
The 20M particles simulations needed 2 minutes per step on the GS320.
The
total energy drifts by no more than 1 per cent, and total angular momentum
is conserved to typically within 0.1 per cent. The results
discussed below are for the high resolution simulation unless
otherwise noted.

\section{Bar Formation and Structural Evolution}

Although these models are formally in equilibrium they are
strongly unstable.  A bar develops by $t$=26 (686 Myr),
and is sustained for the duration of the run.
Visually, the bar grows to its greatest extent at around $t$=50 (1.32
Gyr). The bar then bends and buckles between $t$=70 and $t$=90 (a period
of 526 Myr) to leave a slightly slower bar of comparable length.
There are isophotal twists seen during the bar buckling phase, but
as noted in Shaw et al. (1993) these disappear with the increased
heating of the disc seen during the bar evolution.  After this
short phase, the bar shows elliptical isophotes 
with a butterfly- or dumbbell-shape 
when viewed face-on.   
Fig. 3 shows the face-on views for $t$=50 to 500.
\begin{figure*}
\centering \epsfig{figure=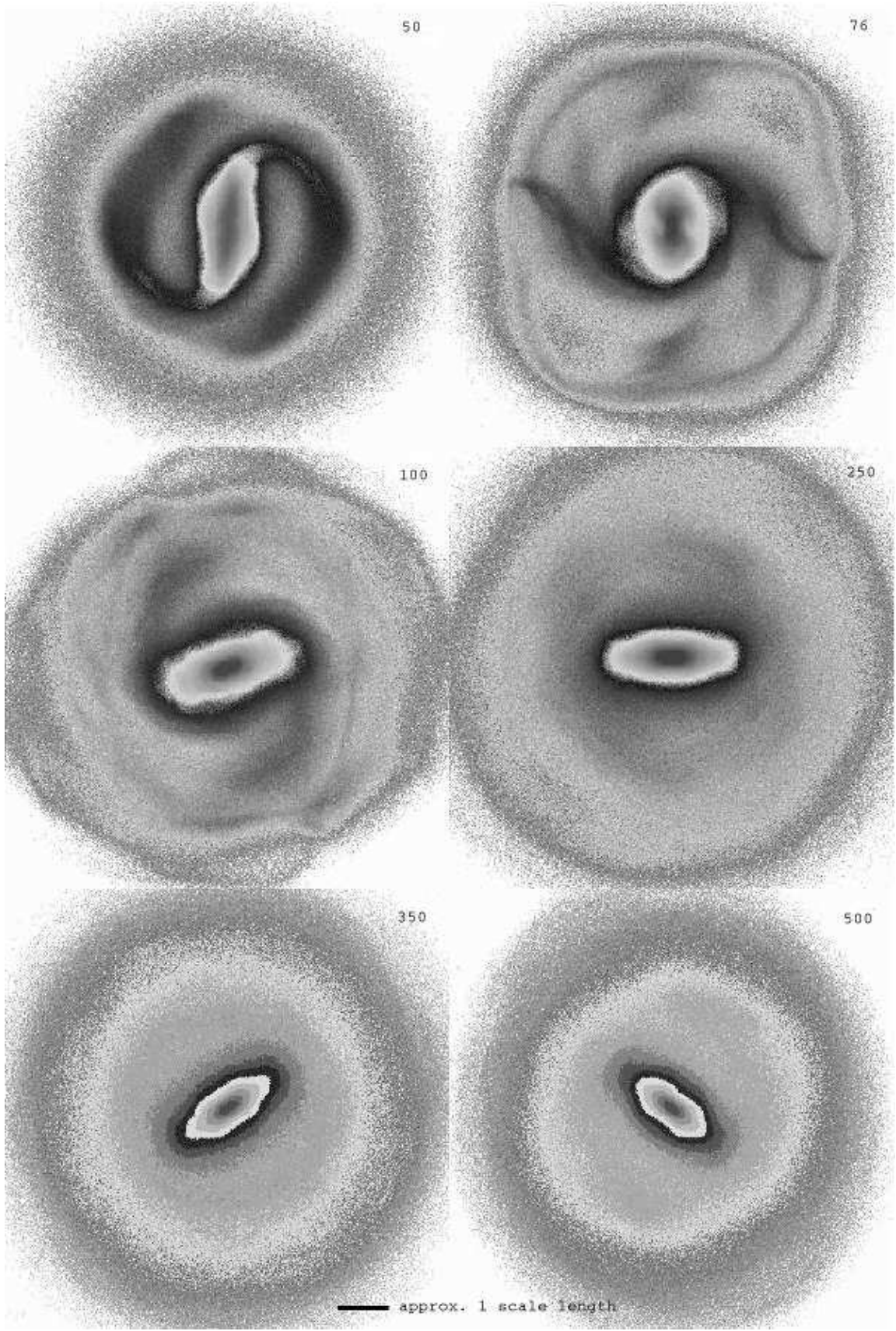,width=6.0in,angle=0}
\caption{Logscale density diagrams of the face-on views of the simulated galaxy at $t$=50, 76, 100, 250, 350, and 500.}
\end{figure*}

The evolution of the bar viewed edge-on clearly reveals the
buckling instability. Initially the disc is thin; the
formation and motion of the bar heat the disc vertically
and azimuthally. At the onset of buckling there is a very short
lived phase of about 260 Myr in which the bar bends into an arc before
buckling into a boxy shape object (edge-on views are  shown
in Figs 4 and 5).  The boxy-shaped bulge remains after buckling,
and slowly evolves into the familiar
peanut shape, which is
most prominent when the bar is viewed edge-on, with its long side in
the plane of the sky. When viewed end-on, the galaxy resembles a
disc with a spherical bulge.
\begin{figure*}
\centering \epsfig{figure=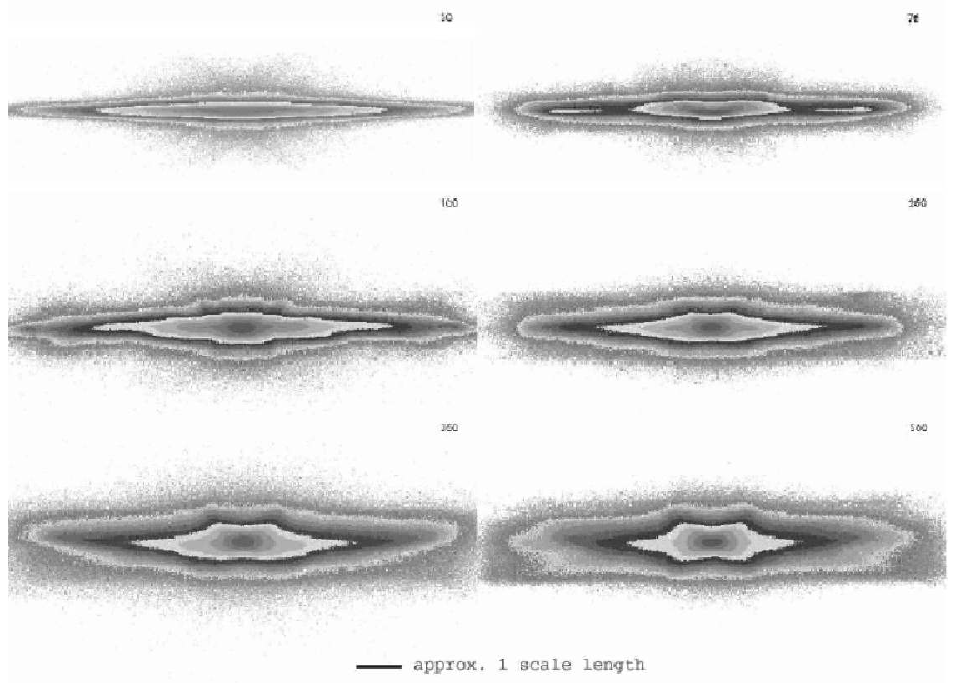,width=6.0in,angle=0}
\caption{Logscale density diagrams of the edge-on views of the simulated galaxy at $t$=50, 76, 100, 250, 350, and 500.}
\end{figure*}
\begin{figure*}
\centering \epsfig{figure=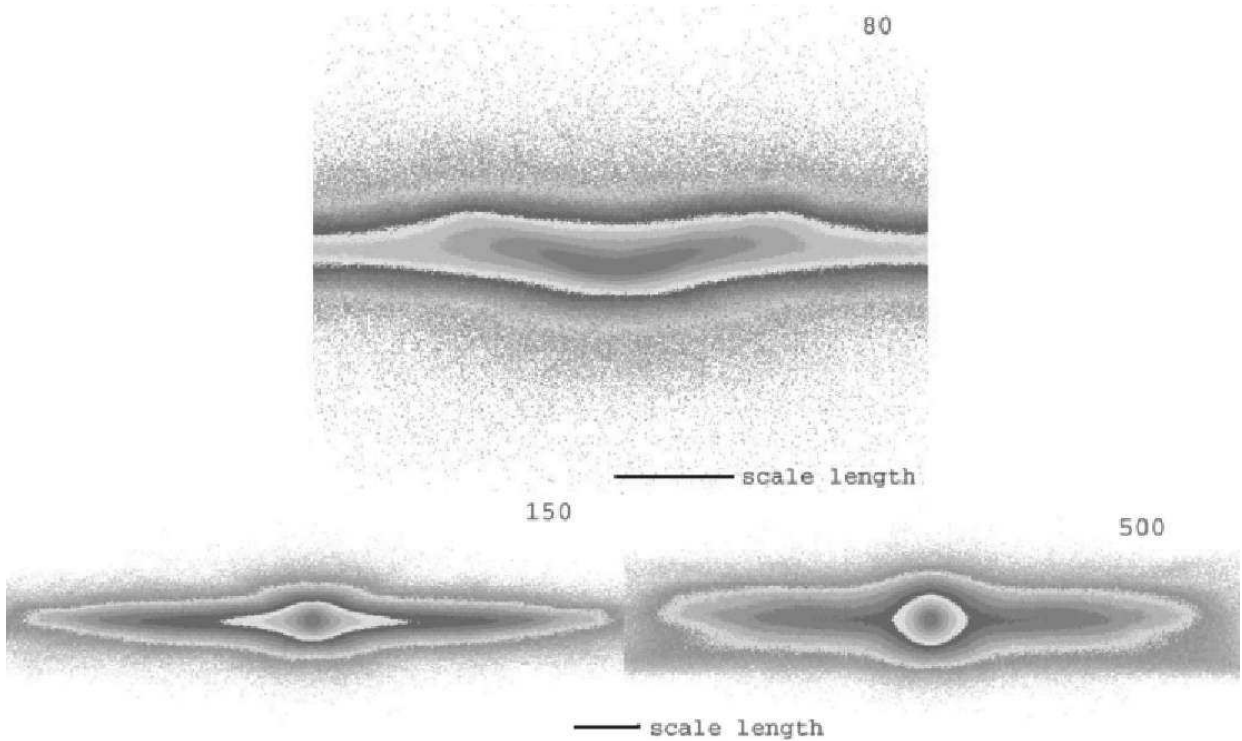,width=6.5in,angle=0}
\caption{The edge-on views of the simulated galaxy showing the warp at $t$=80, and the end on views of the bar at $t$=150 and 500.}
\end{figure*}
The bar remains an obvious component of the disc and shows no sign of
collapsing into a spherical bulge.  There is an animation of both face-on 
and edge-on views available at www.astro.utoronto.ca/~oneill.  

There is an increase in central density following the 
initiation of the bar, which influences the initial slope of the
rotation curve.  Fig. 6 shows the rotation curve at $t$=0, 100,
200, and 400, and it shows the increased slope and peak during the
simulation.  The surface density along the bar is flat
at small $R$ for $t$=50, but rapidly turns exponential following
buckling.  These data will be discussed in more detail
in the following sections.
\begin{figure}
\centering \epsfig{figure=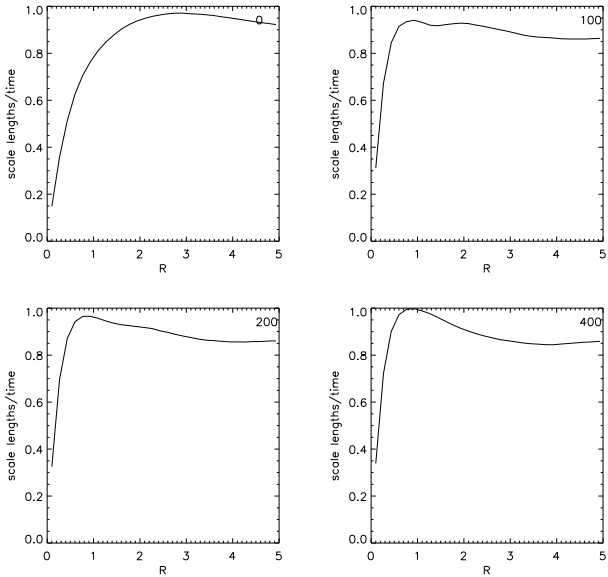,width=8cm,angle=0}
\caption{Rotation curve of the simulated galaxy at $t$=0, 100, 200, and 400.  
The bar causes an increase in central mass, which increases the initial 
slope of the rotation curve as well as raising the maximum value.}
\end{figure}

Even without the gaseous component, this simulated galaxy can be
compared with observations, and some obvious features stand out.  The
face-on views look very much like observed bars, and the edge-on views
are similar to several studies comparing peanut-shaped bulges and
bars (K\&M; Merrifield \& Kuijken 1999).
More detailed comparison with the latter follows in Section 5. 

Recently, A\&M conducted
a series of tests to compare N-body simulations of different mass
models, which are ideally suited to comparison with our results here.
They simulated three
different models:  one disc-dominated (MD), which is similar to ours here,
another
disc-dominated model with an added bulge (MDB), and one halo-dominated
(MH):  all
created using the Hernquist (1993) models.  

Our work agrees quite well with A\&M on isophote shape and rotation 
curve progression
for the first half of the simulations, which span approximately the same time.
Our evolution here disagrees to some extent with their disc dominated
model for the latter half of the run:  the mass build up at the
centre of the disc causes the inner slope of the rotation curve to
become more steep (albeit slightly) and the maximum velocity to increase
even during the final stages; whereas, A\&M found little difference
between the rotation curves for the last quarter of their run.  This
also carries over to the bar isophote structure, as our increasing
central density causes the bar shape to become similar to their
MDB model.  Therefore, the latter half of our
simulation is more readily compared to dumbbell-shaped, face-on and
peanut-shaped, edge-on isophotes, which they produce with the MDB model.
Our low-resolution model, with a similar number of disc particles as were
used by A\&M, shows only boxy isophotes, but was run for half
as long as the high-resolution model.  Since the buckling stage was 
later in this
simulation, the evolution into a peanut-shaped bulge would be
expected after the end of the simulation, so it is uncertain
whether we would agree with A\&M in this case or not.  The rotation curve for
our low-resolution model does not change slope for the last 100 simulation
time units, which may suggest that the mass redistribution does not
continue as long as the high-resolution model run.  

We also investigated the evolution of the halo.
Unlike Weinberg \& Katz (2002), the central density
distribution of our halo remained virtually unchanged.  There were 
several differences
between our models which may account for this. (1) Their bars were much
stronger than ours (30 per cent of the disk mass), appearing earlier in 
the formation of the galaxy out of
a cold, gaseous disc (ours formed in the stellar disc of a galaxy 
initially in equilibrium), which will
have a greater dynamical friction effect on the halo.  (2) Our halo is based
on the King model, which according to Weinberg \& Katz (2002) is 
missing the key resonance for
redistributing the halo mass.  However, the KD models are composite DFs and 
the resulting halo has a mild cusp with 
an effective power law slope of -0.7 at the centre, which is similar to an  
NFW profile.  (3) They 
stress the need for high
resolution, using 4M particles in the halo; they are concerned that tree
codes induce more small scale noise and require higher resolution --
they did not outline how much more, but the 10M particles in our halo should be 
enough to see some indication of halo density redistribution.

\section{Bar Pattern Speed}
T\&W describe the only method available to measure pattern speeds 
of real galaxies from observable quantities alone.
This method uses luminosity, line-of-sight velocity, and tilt angle to
determine the rotation rate of the bar.  To date there have been a few
early-type galaxy bars measured this way, and every one of them has been
found to be fast, with the corotation distance to bar length
ratio between 1 and 1.5.
This result has been difficult to reproduce in simulations, as all the bars 
are greatly slowed due to dynamical friction between the bar and the halo.  
The amount of mass in the halo
compared with the disc is still a point of contention, because the existence
of a flat rotation curve constrains, but does not pinpoint, this ratio.
Studies on this relation refer to `maximal' or `sub-maximal' discs,
where the former has the largest disc to halo mass ratio in the inner
regions of the disc that is consistent with observed rotation curves 
(van Albada \& Sancisi 1986).
D\&S used the dynamical friction
between the bar and halo to show that only a maximal disc
is likely in barred
galaxies, since these bars slow down the least.  The final state of their
maximal disc simulation was nearly a fast bar, with a corotation
to bar length ratio of $1.6 \pm 0.3$.

We used two different methods to
determine the pattern speed of the bar ($\Omega\rm _p$).  The first 
was a direct measurement of bar rotation by determining the orientation angle
of the bar as a function of time.  Fig. 7 is a plot of pattern
speed versus time in the original
simulation units.  Early on, the pattern speed is 0.44 at $t$=26 but
there is a
rapid reduction during the initial bar growth, and by $t$=50 it has slowed
to 0.35, then there is another drop during the buckling phase from $t$=65
to 95, with the pattern speed settling to 0.33 (the following section 
compares these with observations).
After that initial phase of rapid evolution, the pattern speed 
is influenced by
dynamical friction, as it gradually slows to less than 0.21 by the end 
of the simulation.  Even after approximately a Hubble time, it does not reach
a steady state; it is still transferring angular momentum to the halo (see
Fig. 8).
We note in Fig. 7 that the pattern speed
shows significant scatter early on, during bar formation and
buckling phases.  After formation, the pattern speed smoothly, but
quickly, decreases.  During buckling, the scatter is
probably due to difficulty in
precisely measuring bar position in the midst of twisted isophotes.
There is also a series of oscillations which continue after the
buckling phase which we believe to be
real, whose period corresponds to slightly more than the circular
period of the disc at corotation.  We are uncertain at this time
of the cause of these oscillations.
\begin{figure}
\centering \epsfig{figure=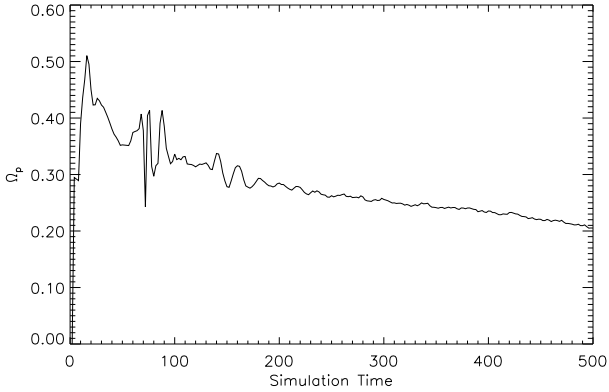,width=8cm,angle=0}
\caption{Pattern speed of the bar vs. simulation time units 
($\rm rad\ t^{-1}$):  the pattern speed continues to slow even after a 
Hubble time.}
\end{figure}
\begin{figure}
\centering \epsfig{figure=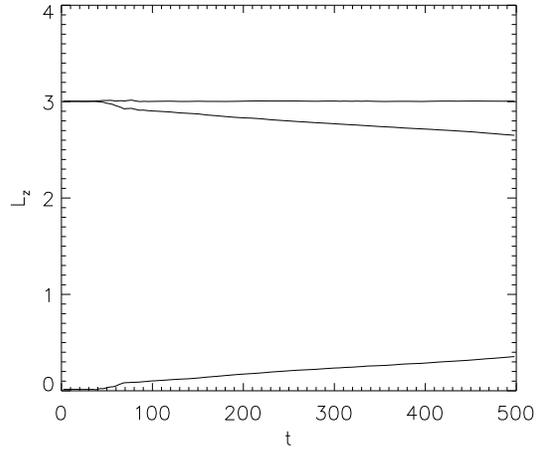,width=7cm}
\caption{Angular momentum of the disk (middle), halo (bottom), and total (top).  The total angular momentum is well conserved and there is a transfer of 10\% of the disk angular momentum to the halo.}
\end{figure}

\begin{figure}
\centering \epsfig{figure=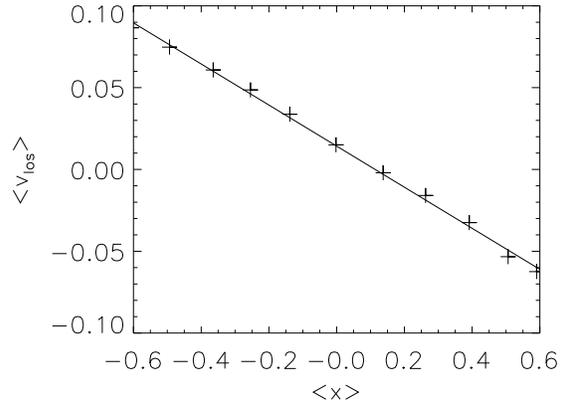,width=8cm,angle=0}
\caption{Determining pattern speed (slope) from luminosity-weighted velocity over position.}
\end{figure}

In order to compare these results with both observations and other
simulations, they must be scaled to comparable units.  
We have taken the R (scale length) divided by v (circular
velocity) to determine a scaling factor (DS subscripts for 
Debattista \& Sellwood, OD subscripts for O'Neill \& Dubinski):   
\begin{equation}
\frac{\Omega_{pDS}}{\Omega_{pOD}} = \frac{v_{DS}}{v_{OD}}\times \frac{R_{OD}}{R_{DS}} \ .
\end{equation}
In comparison with the maximal disc simulation (run 68) from D\&S  
our scaled pattern
speed is slightly lower than theirs for the duration of the
simulation,  $0.086 \rm{\:rad\: t^{-1}}$ instead of about $0.1$ 
after the bar is fully formed, and $0.054 \rm{\:rad\: t^{-1}}$ instead 
of around $0.057$
at the end of our simulation.  Although certainly disc dominated, our
mass model would be more
correctly compared with D\&S control run, which has a similar
disc to halo velocity ratio (their $\eta$ parameter).  
Here, our bar pattern speed 
is slower initially ($0.093 \rm{\:rad\: t^{-1}}$
instead of around $0.15$), but does not slow down as quickly or as
much as their simulation (final speed $0.058 \rm{\:rad\: t^{-1}}$
instead of around $0.041$).  Because we 
are using different simulation techniques and mass models, the 
agreement here is promising, although it should be noted 
that our pattern speed degradation is consistently lower than D\&S.  
This agrees with Athanassoula's (2002b) assertion that bars slow 
down at a faster rate in colder disks.

\begin{figure*}
\centering \epsfig{figure=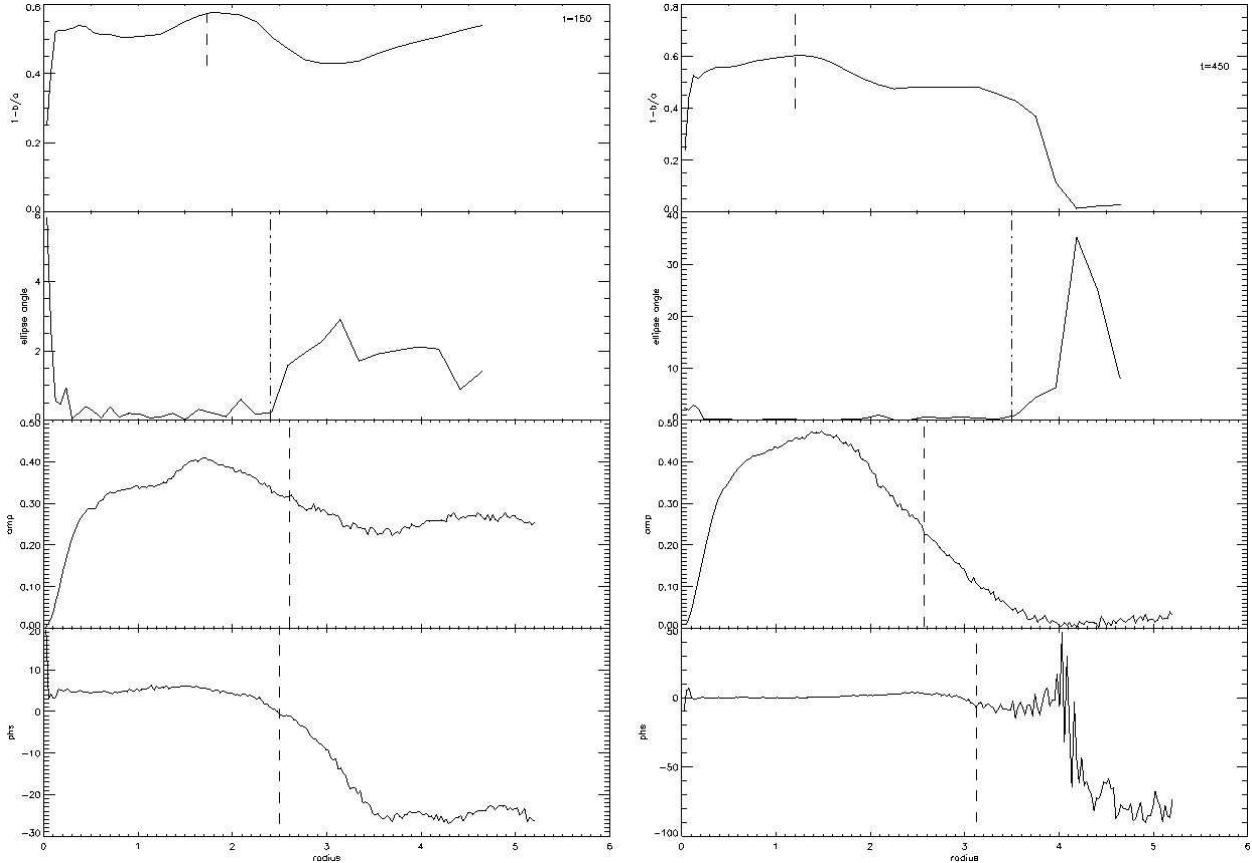,width=6.5in,angle=0}
\caption{Four methods of determining bar length: ellipticity of the bar ($1-b/a$) - the point where 1-$b/a$ is a maximum corresponds to the end of the bar; phase angle of the ellipses - where the angle deviates $5\deg$; m=2 Fourier amplitude and phase; the bar end is measured halfway down the slope on the amplitude plot, and where the phase plot deviates by $5\deg$.  $t$=150 and $t$=450 are shown with the dashed line indicating the measured bar length for each method.}
\end{figure*}

We also measured the bar pattern speeds using the observational
technique described by T\&W.
Taking observations from multiple slits laid
parallel to the apparent major axis of the tilted disc
and finding the luminosity-weighted position and average line of
sight velocities determines the pattern speed:
\begin{equation}
\Omega_p \sin(i) = \frac{<\overline{v_{los}}> - v_o}{<x> - x_o}
\end{equation}
(M\&K).  This method, although direct, is not simple to execute 
observationally.  Since the
basis of this relation is the continuity equation, common
observations using gaseous emissions are disqualified, as the gas cycle
through stars is inherently discontinuous.  The method is therefore
most accurate for those early-type barred spiral
galaxies with the least amount of gas and star formation.  There
are also preferred tilt angles to the galaxy to minimize error
and difficulties surrounding determining the line-of-sight
velocities. It does, however, remove many of the uncertainties
inherent in the indirect methods.

Conversely, this technique is relatively simple to perform on
simulated data.  At
various points during the simulation, snapshots were rotated so
the bar was at a $45^{\circ}$ angle to the x-axis, then tilted
about this axis by $30^{\circ}$, similar to the observed galaxies
listed below. Slits were laid through the centre of the nucleus
and at $\pm$ 1 scale length, parallel to the major axis.  
Fig. 9 plots the luminosity weighted velocity 
versus position, with the assumption of a constant $M/L$ ratio.

This method works very well with the simulated galaxy, with speeds
agreeing to within around 2 per cent of actual for the duration of
the simulation when using the known $30^\circ$ inclination of the
disc. However, in observational analysis the tilt of the galaxy is
not known, and these simulations show that the outer isophotes of
face-on systems are not exactly circular.  Therefore, determining
the tilt of the galaxy by assuming this regular shape will lead to
errors.  Based on the distance to NGC 4596, the outermost detectable
isophote is at 5.8 scale lengths.  At t=200 (5.4 Gyr),
the axis ratio $b/a$ at this radius is 0.72.  If assumed circular,
the computed
tilt angle would be $44.2^\circ$ instead of $30^\circ$, which
would result in a 26 per cent error on the pattern speed as opposed to
the 2.6 per cent error found with the correct inclination. These
errors systematically underestimate the pattern speed  
and are reduced as time progresses
and the outer isophotes become more regular with the disappearance
of any spiral pattern or time-dependent instabilities.

The definition of a fast or slow bar depends on the length of the
bar and the positions of the resonances.  The former is a point of
contention.  A\&M have shown that any one method of determining bar 
length will not work equally well for the duration of the galaxy's lifetime.  
In this paper we use several methods to determine the length of the bar:  
the radius where the phase angle of the elliptical isophotes twists, 
the m=2 component of the Fourier 
amplitude and the m=2 phase component.  The latter two are used by D\&S, and 
all three are used observationally by Aguerri et al. (2003).  Originally we 
had attempted to use Abraham et al.'s
(1999) description for finding the bar length by
choosing the elliptical isophote with the smallest $b/a$ axis
ratio (option (i) or $L_{b/a}$ 
in A\&M's comparison of bar length determination methods), but we found this method 
consistently underrepresented the bar in comparison to the other methods.   
A large drop in this measure during the last third of the simulation causes the 
difference to jump from around 25 per cent to 60 per cent shorter, although we 
find similar ellipticity plots as A\&M for their MD model, where the $1-b/a$ 
measure reaches a peak before the end of the bar, then gradually declines.  The 
peak occurs at a smaller radius for the later time steps.   
The first three methods and the resulting 
bar length determination from each are plotted in Fig. 10.
Fig. 11 shows the average bar length using the first three methods as a 
function of time, with error bars corresponding to the maximum and minimum 
lengths calculated.  The bar length appears to be constant or slightly increasing 
with time.  Also plotted are the corotation to bar length ratios, which are all
under 1.5, indicating the bar is fast throughout the galaxy's evolution.  
D\&S quote a final corotation to bar length ratio
($D_L:a_b$) of $1.57 \pm 0.27$ which they call `acceptably fast
... only barely so'.  There is some agreement between our results and theirs, 
especially if we only take the Fourier component methods into account, 
where we find a final ratio of $1.44 \pm 0.1$.   

\subsection{Comparison with observations}

The actual pattern speed of the bar is scaled to the dimensions of real galaxies 
and compared with observed pattern speeds listed in Table 1.

\begin{table*}
\caption{ Comparison of simulated and observed pattern speeds.}
 \begin{tabular}{lcccccc}
{\bf Galaxy} & {\bf $\rm{R_e}$} & {\bf Velocity} & {\bf Obs. $\Omega_p$} & {\bf Sim $\Omega_p$} (t=26) & {\bf Sim $\Omega_p$}* (t=110) & {\bf Sim $\Omega_p$}* (t=500)\\
       &  (kpc) & ($\rm{km\ s^{-1}}$) & ($\rm{km\ s^{-1}\ kpc^{-1}}$) & ($\rm{km\ s^{-1}\ kpc^{-1}}$) & ($\rm{km\ s^{-1}\ kpc^{-1}}$) & ($\rm{km\ s^{-1}\ kpc^{-1}}$) \\
\hline
NGC 4596 (SBa)$^a$ & $3.2^b$  & $120^b$ & $52\pm13$ & 16.4 & 27.38 & 21.6 \\
NGC 936 (SB0)$^c$ & $3.5^d$ & $280^d$ & $60\pm14$ & 35 & 58.4 & 46 \\
NGC 1023 (SB0)$^e$ & 2.9 & 270 & $87\pm30$ & 40.7 & 68 & 53.6 \\
ESO 139-G009 (SAB0)$^f$ & $4.1^g$ & 314 & $61\pm17$ & 33 & 55 & 44 \\
IC 874 (SB0)$^f$ & $1.88^g$ & 187 & $41.6\pm14.3$ & 43 & 73 & 57 \\
NGC 1308 (SB0)$^f$ & $3.6^g$ & 347 & $99.4\pm34.8$ & 43 & 71 & 56 \\
NGC 1440 (SB0)$^f$ & $1.7^g$ & 283 & $83\pm10$ & 74 & 124 & 98 \\
NGC 3412 (SB0)$^f$ & $2.2^g$ & 205 & $57\pm16$ & 42 & 69 & 55 \\   
\hline
\multicolumn{5}{l}{*Determined with instantaneous scale length (at t=110, $R_e$=2.3 and at t=500, $R_e$=2.8)}\\
\multicolumn{5}{l}{$^a$ Gerssen et al. 1999}\\
\multicolumn{5}{l}{$^b$ Galaxy properties from Kent 1990}\\
\multicolumn{5}{l}{$^c$ Merrifield \& Kuijken 1995}\\
\multicolumn{5}{l}{$^d$ Galaxy properties from Kent 1989}\\
\multicolumn{5}{l}{$^e$ Debattista et al. 2002}\\
\multicolumn{5}{l}{$^f$ Aguerri et al. 2003}\\
\multicolumn{5}{l}{$^f$ Debattista, private communication}\\

\end{tabular}
\end{table*}

The T\&W method directly measures pattern speed from observables,
and is thus more accurate than the other, indirect, methods.  It has 
thus far only been applied to a handful of galaxies, most of which 
are listed in Table 1.  The T\&W method depends on continuity, 
which the life-cycle of gas inherently contradicts, so reliable use 
of this measurement is restricted to early-type galaxies.  

The scaled pattern speeds found in this simulation are 
initially too slow for all but one galaxy (although two more 
are just barely within the lower bound of the observed error bars).  
However, even though the pattern speed decreases with time, the 
redistribution of the disk particles by the bar causes the 
scale length of the disk to increase with time as well.  Measured with 
a double exponential fit to the surface density profile, this increase in 
the disk scale length  
is fairly rapid initially, jumping to 2.3 from 1.0 between $t=50$ and 
$t=110$, and slowly increasing to 2.8 by the end of the simulation.  
This increase more than offsets the pattern speed slow-down, with 
only one comparison still too low at $t=110$ while two are higher than 
observed, and the rest are within the accepted ranges.  By the 
end of the simulation three are again too low, only one is too 
high, and the rest are acceptable (with two at the lower limit of 
the error bars, and two well within the measured range).  
Therefore, for a considerable duration, these long-lived bar is 
rotating at speeds comparable to those observed.  This reiterates 
the corotation to bar length ration finding above: both the simulated 
and observed bars  
are "fast" (ratio less than 1.5).

\begin{figure}
\centering \epsfig{figure=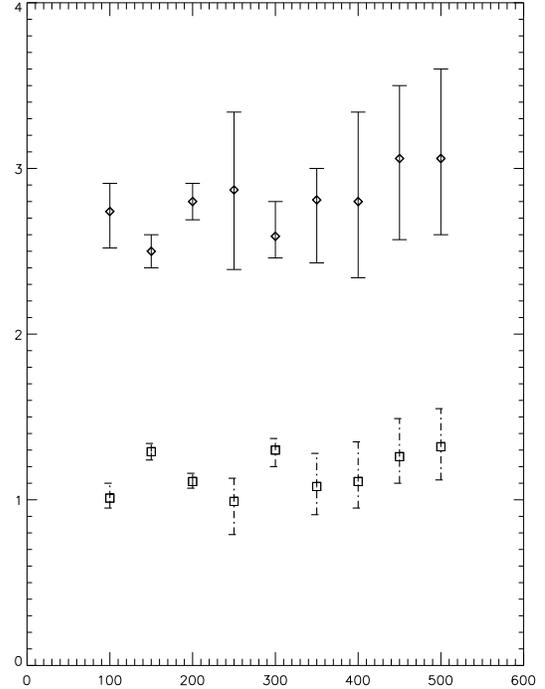,width=8cm,angle=0}
\caption{Bar length (triangles) determined by the average of three techniques plotted in Fig. 10 (ellipse angle, m=2 Fourier amplitude and phase; ellipticity was found not to be a good indicator in this case).  Corotation to bar length ratio is noted with squares at the bottom of the plot; the ratio is always below 1.5, indicating a fast bar.}
\end{figure}

\section{Edge-on Kinematics}

Some attention has recently been placed on the
identification of edge-on barred galaxies.  K\&M, 
Athanassoula \& Bureau (1999) (A\&B), and
B\&A have studied the line-of-sight velocity profile
of peanut-shaped, edge-on galaxies and have shown that their
kinematics are explained by the presence of a bar.  The
characteristic plot of position versus velocity gives a unique
figure-eight pattern for barred galaxies, since some orbits are
depleted in these systems.  Our simulation results have been similarly 
plotted.  After
initial bar formation, the plot appears similar to the unbarred
plots; as the bar develops, the central positions darken while the
density increases, and the areas immediately to the outside of
this central pole suffer some depopulation, similar to the
observational plots in K\&M and A\&B.  With bar buckling, though,
comes increased scatter in the plot, removing any forming pattern, 
and once again leaving a plot which would be considered
unbarred if viewed observationally.  Plots at $t$=50 and $t$=76 are
shown in Fig. 12.
\begin{figure*}
\centering \epsfig{figure=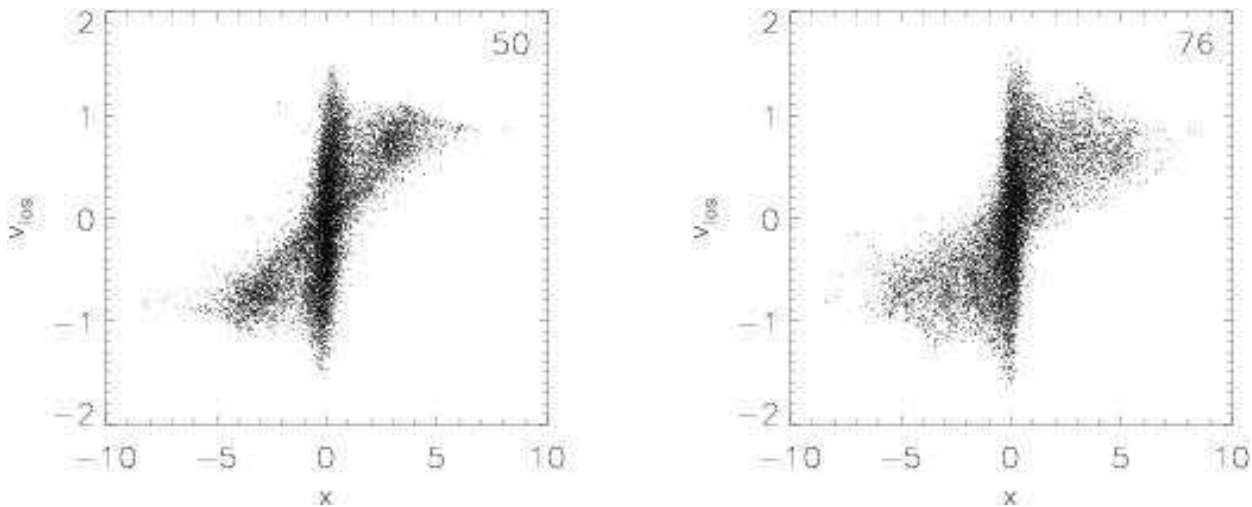,width=6.5in,angle=0}
\caption{Line-of-sight velocity distributions from t=50 and t=76.  A slight figure-eight pattern is emerging at t=50, but is wiped out by the buckling phase at t=76.}
\end{figure*}

The original simulation is composed only of a disc and halo and
was constructed to be in equilibrium with a Toomre Q greater than
1.  This generates a higher velocity dispersion than would be
observed: 0.23 (times maximum velocity) tangential and 0.215
radial dispersion at 2.4 scale lengths.  The radial dispersion
increases to 0.258 by $t$=76. This blurs the figure-eight
pattern in the K\&M plot.  Since all the
observational results used cool gas to determine velocities, our
results would be much closer to the analytically predicted orbits,
and the distinctive figure-eight plot. 

\section{Bulge Model}
To counter part of the possible velocity dispersion problem stated
above, a model
containing a small, compact bulge was constructed in the lower
resolution range (500K particles), keeping a similar halo and
rotation curve (see Section 2 for a more complete description).  The evolution 
of the disc is similar to the
earlier model in that the bar is initiated a little before $t$=20
(540 Myr); it differs in that the initial bar is about 20 per cent smaller,
with dumbbell-shaped inner isophotes when viewed face-on.  
The bar remains smaller than the  
the previous model, by about the same amount for the duration 
of the simulation.   
This is in direct contrast with the findings of A\&M
who found their bulge model to have a longer
bar, although the shape is similar.  However, their bulge is more 
massive in comparison to the disk, and more extended than ours.
Fig. 13 shows $t$=26 to 250.
The initial velocity dispersion (0.146 versus
0.23 above for tangential and 0.18 versus 0.215 for radial) was lower
than our original simulation; however, the line-of-sight velocity
distribution (LOSVD) plots remained similar.  The high resolution
simulation showed much more detail and followed trends toward the
figure-eight shape not seen in the lower resolution plots for the
same initial conditions.  Creating a simulation of the bulge mass
model with higher resolution, with its lowered velocity dispersion,
should yield
clearer LOSVD plots than the original high resolution run, therefore 
we cannot make any conclusions as to whether or not the inflated  
velocity dispersion in the original simulation 
is responsible for masking the figure-eight pattern.
\begin{figure*}
\centering \epsfig{figure=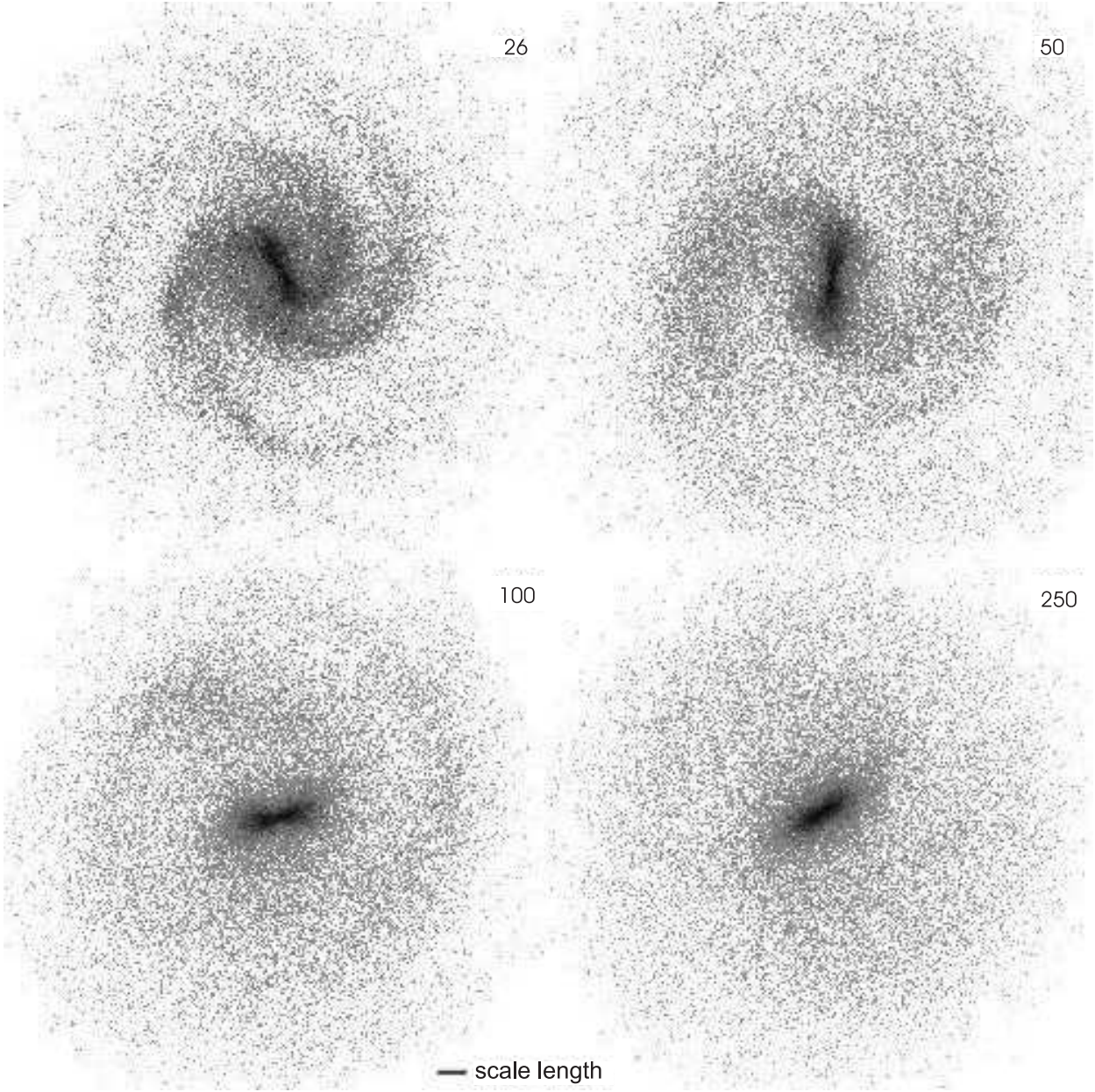,width=6.5in,angle=0}
\caption{Non-contoured logscale face-on views of the bulge model for t=26, 50, 100, and 250.  Only the disk particles are shown.}
\end{figure*}

The pattern speed of the bulge model is higher than the original
for the duration of the simulation, is reasonably
steady after $t$=100 and can be compared with observations,
as earlier.  The result is that the pattern speed scales
up to an acceptable rate for NGC 1023, ISO 139-G009, and NGC 1308, is 
too low for NGC 4596, and too high for NGC 936, IC 874, NGC 1440, and 
NGC 3412 for the whole run.  The final measured  
pattern speed for the bulge model (at $t$=250) is the same as that
for $t$=50 in the original disc+halo model.  Some speed increase is
to be expected with the higher central density, as was mentioned
in D\&S; however, too much central density means the disc is no
longer unstable to bar formation.  The bar was even closer to 
corotation, with a $D_L/a_B$ of near one for most of the run, 
ending with $1.3 \pm ^{0.33}_{0.17}$ 

\section{Halo Bar}

A slightly elongated shape was excited in our halo, with an axis
ratio of around 0.88.  If the alignment of this halo bar were to
coincide with the disc bar, the dynamical friction should be
reduced.  Although the $\Omega_p$ deceleration slows after about
$t$=150, suggesting a decrease in torque on the bar, the halo bar
takes much longer to line up with the disc bar.  The bar
orientations are mostly within $10^{\circ}$ of one another after
t=450, although not locked in at the same relative angle. D\&S
also found an m=2 response in the halo which slowly aligns with
the disc bar.

The pattern speed degradation during the later time steps may not
be solely due to the halo friction.  Misiriotis \& Athanassoula
(2000) have shown that the $\Omega_p$ is slower in thicker discs.
As the bar buckles and the simulation progresses, the scale height
of the disc increases.  This should then also slow the bar, even
if the halo bar is locked into synchronous rotation with the disc
bar, and the dynamical friction between the disc and halo is at a minimum.

\section{Comparison of Early and Late Type Bars}

Elmegreen \& Elmegreen (1985) studied the kinematic and dynamic
properties of different Hubble-types.  They found distinct
photometric differences between early (SB0-SBbc) and late-type galaxies.
Early-type galaxy bars appear to end at or near corotation and are twice
the length of the turnover radius of the rotation curve.  Their
luminosity profile is flat compared with the exponential decrease of
the rest of the disc.  Late-type bars are shorter, seem 
to end near the inner Lindblad resonance, and are shorter
than the turnover radius.  They have an exponential luminosity
profile with a different slope than that of the disc.
\begin{figure}
\centering \epsfig{figure=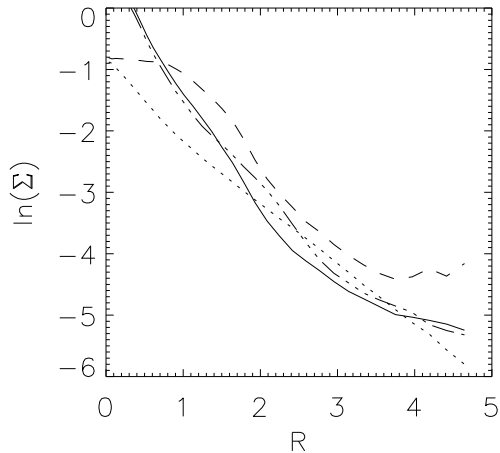,width=8cm}
\caption{Surface density (luminosity) for t=0 (dotted), t=50 (dashed), t=250 (dash-dot), and t=500 (solid).  t=50 shows an initial flat-topped profile, but the flat section is about half the length of the bar.  The later times show an evolution to a double exponential.}
\end{figure}

The properties of the bar evolves during our simulation.  By assuming 
a constant $M/L$, the density profile of the bar can be taken as 
representative of a luminosity profile (Fig. 14).  Soon after the 
bar forms, a small section of the bar near the core has a flat density 
profile, 
with a near exponential decrease for the 
remainder of the bar.  Later time steps show only exponential profiles.  The 
bar length is near the corotation resonance for the duration of the run, 
and the bars are always longer than the turnover radius and the inner 
Lindblad resonance.  Fig. 15 contains plots of the Lindblad resonances 
for the bar: $\Omega$, $\Omega-\kappa/2$, and $\Omega+\kappa/2$, which 
are determined based on an azimuthally averaged 
redistribution of the particles to make the disk axisymmetric.  The bar 
length criteria agree with an early-type 
galaxy, but the density profile is late-type.  Another point to note is 
that early-type galaxies are also characterized by a large bulge to disk 
ratio, whereas our models, at most, have a very small bulge.

The main results we are comparing in this study (evolution of pattern speed, 
corotation to bar length ratio) may actually be similar for all barred 
galaxies.  Kinematical studies have shown that bars should be fast even 
in late-type galaxies (e.g. due to dust lane placement 
Athanassoula 1992, and hydrodynamical simulations Weiner et al. 2001).  
Therefore, our small or no bulge galaxy with fast bar with an exponential 
density profile is fully consistent with a real late-type galaxy.  
 
\begin{figure*}
\centering \epsfig{figure=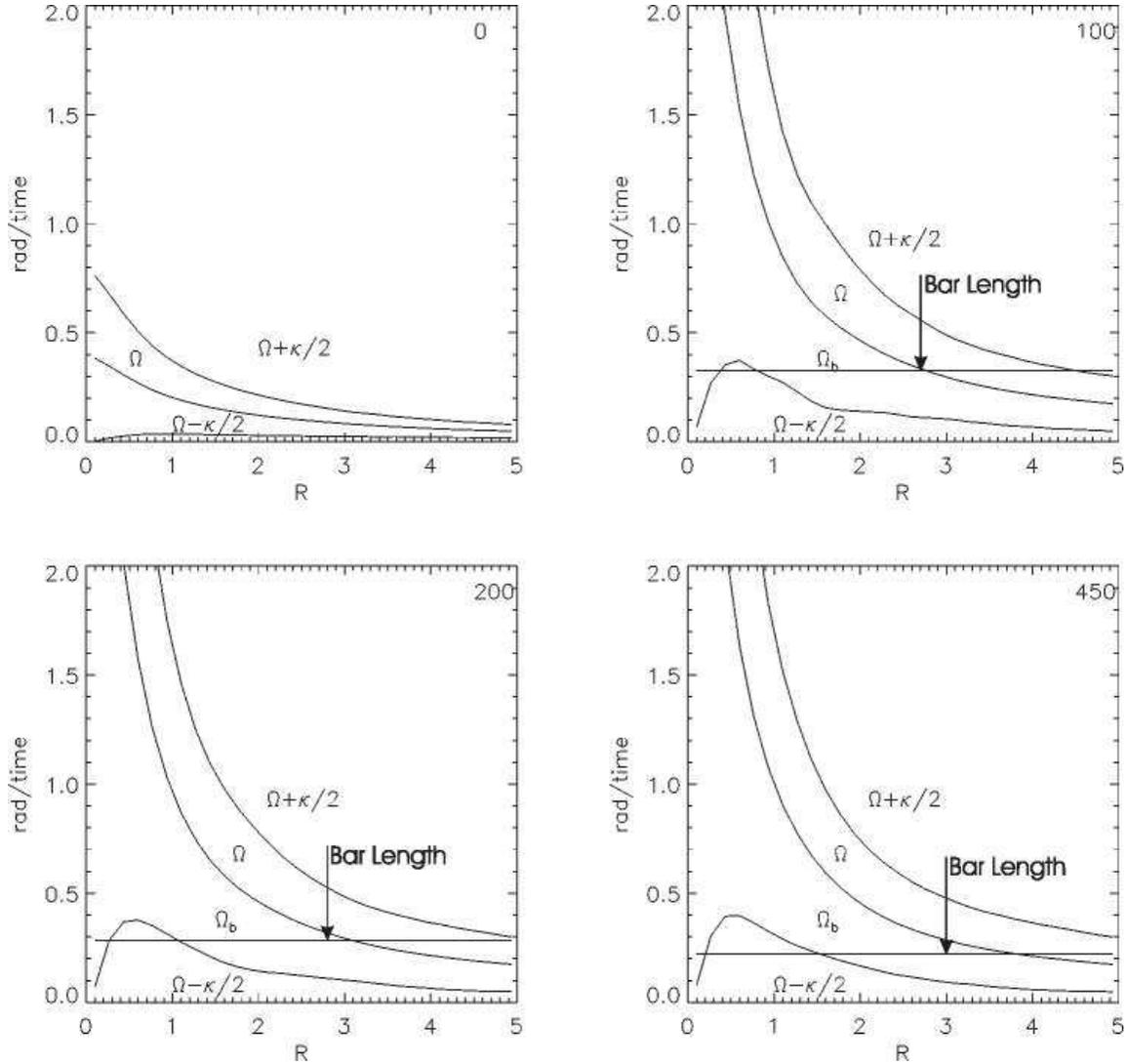,width=6in,angle=0}
\caption{Lindblad resonances and the bar pattern speed for $t$=0, 100, 200, 
and 450.  In all cases the bar is slow; at $t$=450 the bar length is around 
a scale length and within the ILR as predicted for late-type barred spirals 
by Elmegreen and Elmegreen (1985)}
\end{figure*}

\section{Discussion}
\subsection{Resolution}

Along with the 20M particle simulation, a 500K
particle run with the same initial conditions was conducted for comparison.  
The pattern speed is very similar, although after t=350 the 
lower resolution model slows a little more than quoted above ($\Omega\rm_p = 0.18$ 
vs. $0.20$).  The bar lengths, based only on the $m=2$ amplitude and 
phase change, are in general shorter than in the 20M simulation before t=350, 
and longer thereafter, showing a distinct trend to increasing the 
bar length over time.  Most of the values are within the error bars of 
Fig. 11, and there is therefore at least in nominal agreement between the 
two simulations.
 
The high resolution simulation showed much more detail
in the bar.  For example, isophote twists during the formation and
buckling of the bar were clearly seen, whereas they were absent in
the lower resolution model.  The T\&W
method of finding pattern speed was also more accurate with the
larger number of particles.  The 500K model results were within 5-10 per cent 
of the actual pattern speed as opposed to around 2 per cent for the 
high resolution results.  

By the midpoint of the runs, scale height calculations were
comparable with the original simulation, but
at t=100 the average scale height for the high resolution
simulation seems to be higher than the 500K model.  This is most
probably due to the fact that the bar in the high resolution
simulation has already buckled at this point, whereas the low
resolution model has not; buckling heats the disc.
The scale height is of some interest, as it shows the interaction
of the bar with the disc, and an increased scale height has been
shown to decrease the pattern speed (Misiriotis \& Athanassoula
2000).

Therefore, it can be concluded that the lower resolution is
capable of outlining many dynamical properties, but higher
resolution is needed for some detailed comparisons with real
galaxies.

\subsection{Code Comparison}

Victor Debattista and Jerry Sellwood generously offered the use
of their mass model initial conditions to run in our simulation code
for comparison.  Their simulation was run using a 3-D Cartesian
particle-mesh code, with a
grid spacing of $0.2 R\rm _d$, and a total particle count of 600K (102K in
the disc).
Here we run the same particles (run 68 in D\&S, maximum
disc model) with a tree code (Dubinski 1996).  We
initially had problems with our small softening length, until we
changed this to correspond with the grid resolution in their original
simulation (equivalent to a Plummer softening of 0.4 scale lengths
(1.3 kpc) instead of 0.0125 for
the halo and 0.005 for the disc), as their particles are set up in equilibrium
with respect to the mesh.  They ran many different initial conditions
during their research, and we found the $Q$=1.5 model (unpublished) to evolve similarly
to our code.  We found a pattern speed of $0.095\rm{\ rad\
t^{-1}}$ at $t$= 1000 compared
to around 0.07 plotted in their paper (their units), and a shorter 
bar than the original, although when Debattista and 
Sellwood (private communication) ran the same model 
in a higher resolution mesh code this result was corroborated.  The
equilibrium of the particles with the original mesh is most likely the reason
for the disparity.  The higher resolution leads to stronger forces in the
central region of the galaxy, which the initial velocity dispersion does not
support, causing a slight collapse of the disc particles to the centre at
the beginning of the simulation.  This causes the rotation curve to rise more
steeply, which leads to a shorter bar (Combes \& Elmegreen 1993).  
The shorter bar feels less friction from the halo, which results in less 
pattern speed deceleration and the higher pattern speeds found here.

Run 68 in D\&S was actually conducted with a $Q$=0.05 model.  Due to earlier
attempts on our models, we were concerned about the low $Q$ value and 
the stability of the
disc.  We were unsuccessful in reproducing the output from this model,
as we were in our own discs with a $Q$ much below 1.0.  Their results may be
due to the way the particles were set up in relation to their grid, or the
grid method may be supplying some artificial stability to the model.

A copy of our low resolution initial conditions were also forwarded back to
Debattista and Sellwood.  Using a new hybrid grid code with softening of 
$0.025 R_d$ and a time step of 0.025 they found bar lengths between 2.5 and 3, 
well within the error bars of Fig. 11.  They measured the strength of the 
bar by calculating the $m=2$ amplitude of the whole disc as a function of 
time.  The resulting strength is very similar to ours for the same resolution, 
however we find a greater bar strength for the 20M run between $t=150$ and $t=450$.  
The increase may be due to the larger number of particles available for 
angular momentum transfer in that model.  Although we agreed well until $t=150$, 
they found a significantly higher pattern speed thereafter, as their simulation 
did not slow down as much as ours.  The reason for this is unclear at the moment.  

While initial differences were seen, the comparison between these codes 
has shown most of the results to be reproducible.  That said, direct code 
comparison is very much needed in this 
field to determine the influence of method versus initial conditions.

\section{Conclusions and Future Work}

The three N-body galaxy models created here generally concur with previous 
research in structure (face-on and edge-on shapes of bars).  

The bars measured here are all fast, with pattern speeds in the range  
of those observed directly in early-type galaxies.  Although the 
pattern speeds of late-type galaxies, which would be a more appropriate 
comparison to these simulations, cannot be directly measured, kinematical 
studies have agreed that their bars are also fast 
(e.g. van Albada \& Sanders 1982; Hunter et al. 1988; Athanassoula 1992; 
Lindblad et al. 1996; Weiner et al. 2001).

The corotation to bar axis ratio in the high resolution simulation
is less than 1.5, which is designated a fast bar.  This is 
marginally consistent with the D\&S simulations,
although our models did not require a fully maximal disk to remain fast.

A recent preprint by Valenzuela \& Klypin (astro-ph/0204028) outlined 
a simulation which used 780K 
particles (model B), Hernquist initial conditions, and an adaptive 
tree-mesh code that has a higher resolution at the centre than the less 
active outer regions.  Their 
spatial resolution in the central regions is slightly lower than ours, 
and they use many fewer particles than our high resolution model.  
Many of their results 
are in agreement with those presented here: namely, comparable 
angular momentum transfer, the central density build up created 
by the bar and its affect on the rotation curve, the lack of alteration 
of the dark matter density distribution, and the 
double exponential disk density profile.  We disagree  
in bar length 
(theirs are much shorter), actual pattern speed (they find 
a more rapid rotation), and pattern speed degradation (their pattern 
speed is constant).  Our disc to halo mass ratio and Q parameter are 
different, though, which may account for some of the discrepancies. 

By comparing our 20M particle simulation with a 500K 
particle model, it was determined that the basic dynamics
of the system can be reproduced with lower resolution (pattern
speed, overall evolution), but if observational techniques are to
be used on the results, higher resolution is needed.  The
techniques used to varying degrees of success here were the T\&W
method to find pattern speed and the K\&M edge-on
line-of-sight velocity distribution plots to find bar orbits.

The results presented here agree with much of the previous work done 
on isolated, bar unstable galaxy simulations.  We have shown resolution 
is no longer a large source of error.  We suggest further investigations 
should concentrate on generating early-type
barred spiral galaxies with high resolution low softening codes. 
Also, more
observational data of real pattern speeds, especially for
late-type galaxies, would be 
a great asset to the field.

\section{Acknowledgments}

We would like to acknowledge the assistance of Victor Debattista,  
Jerry Sellwood, and the anonymous referee for detailed and 
thought provoking comments, as well as the NSERC and the Canadian 
Foundation for Innovation.

\appendix

\bsp

\label{lastpage}

\end{document}